# Unreliability of two-band model analysis of magnetoresistivities in unveiling temperature-driven Lifshitz transition

Jing Xu[1,2], Yu Wang[3], Samuel E. Pate[1,4], Yanglin Zhu[3], Zhiqiang Mao[3], Xufeng Zhang[2], Xiuquan Zhou[1], Ulrich Welp[1], Wai-Kwong Kwok[1], Duck Young Chung[1], Mercouri G. Kanatzidis[1,5], and Zhi-Li Xiao[1,4,*]

[1]*Materials Science Division, Argonne National Laboratory, Argonne, Illinois 60439, USA*

[2]*Center for Nanoscale Materials, Argonne National Laboratory, Argonne, Illinois 60439, USA*

[3]*Department of Physics, Pennsylvania State University, University Park, Pennsylvania 16802, USA*

[4]*Department of Physics, Northern Illinois University, DeKalb, Illinois 60115, USA*

[5]*Department of Chemistry, Northwestern University, Evanston, Illinois 60208, USA*

Recently, anomalies in the temperature dependences of the carrier density and/or mobility derived from analysis of the magnetoresistivities using the conventional two-band model have been used to unveil intriguing temperature-induced Lifshitz transitions in various materials. For instance, two temperature-driven Lifshitz transitions were inferred to exist in the Dirac nodal-line semimetal ZrSiSe, based on two-band model analysis of the Hall magnetoconductivities where the second band exhibits a change in the carrier type from holes to electrons when the temperature decreases below $T = 106$ K and a dip is observed in the mobility versus temperature curve at $T = 80$ K. Here, we revisit the experiments and two-band model analysis on ZrSiSe. We show that the anomalies in the second band may be spurious, because the first band dominates the Hall magnetoconductivities at $T > 80$ K, making the carrier type and mobility obtained for the second band from the two-band model analysis unreliable. That is, care must be taken in interpreting these anomalies as evidences for temperature-driven Lifshitz transitions. Our skepticism on the existence of such phase transitions in ZrSiSe is further supported by the validation of the Kohler's rule for magnetoresistances for $T \leq 180$ K. This work showcases potential issues in interpreting anomalies in the temperature dependence of the carrier density and mobility derived from the analysis of magnetoconductivities or magnetoresistivities using the conventional two-band model.

## I. INTRODUCTION

As an electronic topological transition without lattice symmetry breaking, the Lifshitz transition, at which the Fermi surface topology changes due to the variation of the Fermi energy and the band structure [1,2] has attracted extensive attention recently [2-24]. It can be induced by various parameters such as pressure and magnetic field in a variety of materials including superconductors as well as topological materials [2-17]. For example, magnetic field induced Lifshitz transitions were observed in UCoGe [2] and CeIrIn$_5$ [5]. Black phosphorus [3] and ZrSiTe [4] exhibit Lifshitz transitions under pressures. Ba(Fe$_{1-x}$Co$_x$)$_2$As$_2$ [6] and V$_{1-x}$Ti$_x$Al$_3$ [7] were found to undergo Lifshitz transitions at the doping levels of $x = 0.038$ and 0.35, respectively. Temperature-induced Lifshitz transitions were reported in WTe$_2$ [8], ZrTe$_5$ [9], ZrSiSe [10], InTe$_{1-\delta}$ [11], Ag$_2$Se [12], Nb$_2$Se$_3$ [13], and ZrSiTe [14]. Surface charge doping can also induce a Lifshitz transition in NbAs [15]. Ultrafast dynamical Lifshitz transitions can even be induced by photons in MoTe$_2$ [16] and ZrSiTe [17].

Angle-resolved photoemission spectroscopy (ARPES), which can probe direct information on the Fermi surface and electronic structure of materials, has been the most common experimental technique in revealing Lifshitz transitions [6,8,9,15-18]. de Haas–van Alphen [2] and Shubnikov–de Haas [3,5] quantum oscillations also have been used to investigate Lifshitz transitions by probing the change in the Fermi surface (FS) topology. Methods such as Raman spectroscopy [4] and nuclear magnetic resonance (NMR) spectroscopy [14] also have been used, though they can only detect Lifshitz transition induced anomalies in other properties rather than the changes in the Fermi surface and the band structure. Recently, conventional transport measurements on the longitudinal resistivity $\rho_{xx}$ and/or the Hall resistivity $\rho_{xy}$ [7,10-13,19-24] have increasingly utilized to uncover Lifshitz transitions, since the change in Fermi surface topology can result in anomalies in the evolution of the carrier density $n$ and/or mobility $\mu$, for example, with doping [7,19] or temperature [10-13,20-24].

In this work we intend to show that whereas a Lifshitz transition can induce anomalies in $n$ and $\mu$, the converse may not be necessarily correct. For example, such anomalies could also arise from the change in the shape and size of the Fermi surface in the absence of a Lifshitz transition. Below, we reveal that artificial anomalies may also occur in multiband materials, due to the lack of a reliable determination of the density and mobility valuses obtained fromtransport measurements. Currently, $n$ and $\mu$ are obtained from: (i) the Hall resistivity $\rho_{xy}$ using the single band picture [11,12]; (ii) the two-band model fittings on the Hall magnetoconductivity $\sigma_{xy}$ [10,19-21,24], Hall magnetoresistivity $\rho_{xy}$ [22,23] or simultaneously on both the Hall and longitudinal magnetoconductivities $\sigma_{xx}$ and $\sigma_{xy}$ converted from the measured magnetoresistivities $\rho_{xx}$ and $\rho_{xy}$ [13].

In the first method, for a single-band (or one dominant band) system, the Hall resistivity $\rho_{xy}$ is proportional to the magnetic field $H$, i.e., $\rho_{xy} = R_H H$, where $R_H$ is the Hall coefficient. The carrier density can be calculated using $n_H = 1/(eR_H)$ and the mobility $\mu$ can be derived using $\mu = R_H/\rho_0$, with $\rho_0$ being the longitudinal resistivity at zero-field [11]. As discussed in Ref.25, however, a linear relationship between $\rho_{xy}$ and $H$ can also exist in a two-band or multi-band material with varying mobilities among different bands, as long as $\mu_i H \ll 1$, where $\mu_i$ is the carrier mobility of the $i^{th}$ band. Furthermore, a compensated two-band system can always have a linear dependence of $\rho_{xy}$ on $H$, regardless of the values of the carrier mobilities and the magnetic field. In this case, the value of $n/n_H$ depends on the ratio of the mobilities in the two bands. That is, the values estimated using the single band picture can differ from the carrier density and mobility in the material. Thus, their anomalies such as peaks/discontinuity in the temperature dependence of $n$ and $\mu$ may not reflect Lifshitz transitions.

Here, we tackle the suitability of the second method, i.e., using the anomalies in the evolution of the carrier densities and mobilities derived from two-band model fittings of the experimental magnetoresistivities/magnetoconductivites as evidences for a Lifshitz transition. We revisit the temperature-driven Lifshitz transition recently reported by Chen et al in ZrSiSe [10], using deduction from anomalies in the temperature dependence of the carrier density and mobility derived by analyzing the Hall magnetoconductivity $\sigma_{xy}$ using the two-band model. We conduct the same measurements and data analysis on ZrSiSe and show that the observed anomalies may not represent evidences for a temperature-driven Lifshitz transition. We also find that the magnetoresistivities $\rho_{xx}$ obey the Kohler's rule that is valid only when the carrier density is temperature insensitive [25], casting further doubts on the existence of a temperature-induced Lifshitz transition in ZrSiSe.

## II. MATERIALS AND METHODS

We conducted resistance measurements on two platelike ZrSiSe crystals grown using a chemical vapor transport method [26]. Electrical leads were gold wires glued to the crystal using silver epoxy H20E. The magnetic field is applied along the $c$-axis of the crystal while the current flows in the $ab$ plane, i.e., the magnetic field is always perpendicular to the current. We measured both longitudinal and Hall resistances $R_{xx}(H)$ and $R_{xy}(H)$ curves at various fixed temperatures, enabling the calculation of the magnetoresistivities $\rho_{xx}(H) = R_{xx}wd/l$ and $\rho_{xy}(H) = R_{xy}d$, where $w$, $d$, and $l$ are the width, thickness of the sample and the separation between the voltage contacts, respectively. They were converted into magnetoconductivities using equations $\sigma_{xy} = \rho_{xy}/(\rho_{xx}^2+\rho_{xy}^2)$ and $\sigma_{xx} = \rho_{xx}/(\rho_{xx}^2+\rho_{xy}^2)$ [10]. The magnetoresistance is defined as $MR = [\rho_{xx}(H) -$ $\rho_0]/\rho_0$, where $\rho_{xx}$ and $\rho_0$ are the resistivities at a fixed temperature with and without the presence of a magnetic field, respectively. We used Origin 2021b from OriginLab Corporation [27] in data plotting and analysis including the fittings presented below.

## III. RESULTS AND DISCUSSION

We measured two samples which exhibit similar behavior, with data from one of them presented here. Figures 1(a) and 1(b) show the magnetic field dependences of the longitudinal resistivity $\rho_{xx}(H)$ and Hall resistivity $\rho_{xy}(H)$ at various temperatures, which resemble those reported in Ref.10. For example, the $\rho_{xy}(H)$ curves exhibit the reported three characteristic features, including the increased nonlinearity as the temperature is lowered, positive slope for $T >$ 40 K, and a slope change in the curves for $T \leq 40$ K. The measured $\rho_{xx}(H)$ and $\rho_{xy}(H)$ enable the calculation of the magnetoconductivities, as presented in Figs.2(a) and 2(c) for $\sigma_{xy}$ and Figs.2(b) and 2(d) for $\sigma_{xx}$ for $T = 20$ K and 300 K, respectively (see Fig.S1 for all $\sigma_{xy}$ curves calculated from data in Fig.1, which show the similar trend with temperature as those in Fig.2(c) of Ref.10).

As discussed by Chen et al [10], the slope of the $\rho_{xy}(H)$ curve may provide qualitative information on the relative contributions of the two carrier types. For example, a positive slope may indicate the dominance of the holes in the electronic transport. Quantitatively, they obtained the carrier density and mobility by applying the conventional two-band model on the Hall magnetoconductivity, which can be expressed as

$$\sigma_{xy}(H) = e\left(\frac{n_1\mu_1^2}{1+\mu_1^2 H^2} - \frac{n_2\mu_2^2}{1+\mu_2^2 H^2}\right)H \qquad (1)$$

where $e$ is the charge of the electron, $n_1$ and $n_2$ are the densities and $\mu_1$ and $\mu_2$ are mobilities of the carriers in the first and the second bands, respectively. Note that we apply a minus sign to the second term in Eq.1 so that the derived $n_2$ for the electron band will be positive rather than negative in Ref.10, as required in the calculation of $\sigma_{xx}(H)$ presented below.

By fitting the experimental $\sigma_{xy}(H)$ curves obtained at various temperatures using Eq.1, Chen et al found that the second band exhibits a change in the carrier type from holes at $T > 106$ K to electrons at $T < 106$ K and a dip in the mobility versus temperature curve at $T = 80$ K. Along with other characterizations they concluded that the temperature induces two Lifshitz transitions, at $T = 80$ K and 106 K, respectively.

Here we do not aim to prove that Lifshitz transitions do not exist in ZrSiSe. We are also not against using the two-band model fitting to estimate the carrier density and mobility in two-band [28] and multi-band materials [10,13,19-24]. Instead, we focus on demonstrating that the two-band model fitting, though it limits the free parameters to four, can result in uncertain outcomes for ZrSiSe. We urge cautions to be taken when drawing conclusions on a multi-band system from the carrier density and mobility derived from the two-band model, such as claiming Lifshitz transitions based on anomalies in their temperature dependence.

We followed the same analysis procedure, i.e., using Eq.1 to fit the $\sigma_{xy}(H)$ curves for various temperatures to obtain the carrier density and mobility. As presented in Figs.2(a) and 2(c) for $T = 20$ K and 300 K as examples, Eq.1 can indeed describe the experimental data very well. The derived carrier densities and mobilities at various temperatures for the first and second bands are presented in Figs.2(e) and 2(f), respectively. While both the values and the temperature dependence of the carrier density and mobility for the first band are consistent with those reported in Ref.10, anomalies indeed occur at $T \sim 80$ K in the temperature dependence of the carrier density and mobility of the second band: at $T \leq 80$ K, $n_2$ and $\mu_2$ smoothly decrease with increasing temperature and their values are comparable to those ($n_1$ and $\mu_1$) of the first band. On the other hand, the values of both $n_2$ and $\mu_2$ for $T > 80$ K seem to be random, with $n_2$ being ~1 order of magnitudes larger than those of $n_1$. The randomness of the $n_2$ and $\mu_2$ values indicates that four free parameters in Eq.1 are too many for the fittings, since the fits did not converge, indicating the existence of mutual dependency between parameters [27]. In fact, the values of $\mu_2$ are orders of magnitudes smaller than those of $\mu_1$, resulting in the value ($\sigma_{xy\_2}$) of the second term to be much smaller than that ($\sigma_{xy\_1}$) of the first term in Eq.1. This can be seen clearly in Fig.3, showing that the ratio of $\sigma_{xy\_2}/\sigma_{xy\_1}$ depends on both the temperature and magnetic field. At low temperatures, the second term $\sigma_{xy\_2}$ does contribute significantly to the total value, though it never overwhelms the first term. At high temperatures, the contribution of the second term is negligible, e.g., less than 4% of that of the first term for $T \geq 180$ K. That is, $n_2$ and $\mu_2$ can be any values if $\sigma_{xy\_2}$ is negligible. This implies that Hall magnetoconductivities $\sigma_{xy}(H)$ at high temperatures can be described with only the first term, i.e., reducing the two-band model to one-band model. In Figs.4(b) we present one-band fittings of the $\sigma_{xy}(H)$ curves for various temperatures, showing good overlaps and significant disparities between the fitting curves and experimental data obtained at high ($T$ = 250 K and 100 K) and low ($T$ = 80 K and 20 K) temperatures, respectively. To better reveal the temperature effects on the fitting, we present in Fig.4(a) the temperature dependence of the Adj. R-square, which is a modified version of R-square and is used to quantify how well a model fits the data [29]. A value of less than but closer to 1 corresponds a better fit. It indicates that at $T > 80$ K, the one-band model can describe the experimental $\sigma_{xy}(H)$ data very well while the disparity becomes more pronounced with decreasing temperature for $T \leq 80$ K. This reaffirms that the anomalies (randomness in $n_2$ and $\mu_2$) at $T > 80$ K are caused by too many free parameters in Eq.1 in fitting the experimental data.

We note that the derived $n_2$ and $\mu_2$ in Ref.10 do not exhibit the randomness of those presented in Fig.2(f), though in both cases the anomalies concurrently occur at $T \approx 80$ K and the values $\sigma_{xy\_2}$ of the second term in Ref.10 are also negligible for $T > 100$ K (because $n_2$ is much smaller than $n_1$). It may be caused by the programs used for the fittings or the ways in parameter initialization, i.e., choosing the initial parameters to start the fittings. In any case, the outcomes of a fitting to only $\sigma_{xy}(H)$ with four free parameters need to be scrutinized. For this purpose, we present in Fig.2(b) and 2(d) comparisons of the experimental $\sigma_{xx}(H)$ for $T = 20$ K and 300 K to those calculated using the two-band model

$$\sigma_{xx}(H) = e\left(\frac{n_1\mu_1}{1+\mu_1^2 H^2} + \frac{n_2\mu_2}{1+\mu_2^2 H^2}\right) \quad (2)$$

and the values of $n_1$, $\mu_1$, $n_2$ and $\mu_2$ obtained in the $\sigma_{xy}(H)$ fittings in Fig.2(a) and 2(c). The pronounced disparities clearly show potential issues related to obtaining the carrier density and mobility using a two-band model to fit only the $\sigma_{xy}(H)$ curves. This can be further illustrated by simultaneously fitting both $\sigma_{xx}(H)$ and $\sigma_{xy}(H)$ using Eq.1 and Eq.2 of the two-band model, as presented in Fig.S2. While the fits for $T = 300$ K are reasonably good, the fits for $T = 20$ K show significant disparity, as shown in Figs.S2(a) and S2(b). The derived carrier densities and mobilities presented in Figs.S2(c) and S2(d) show different behavior from those in Fig.2(e) and 2(f). This shows that a Lifshitz transition could occur at a different temperature if it were inferred from an anomaly in the temperature dependence of the carrier density. Thus, anomalies in the temperature dependence of the carrier density and mobility obtained through two-band model analysis of the magnetotransport data do not necessarily denote Lifshitz transitions.

Since ZrSiSe has three bands [26,30], we also analyzed the experimental data using a three-band model. Obviously, this approach cannot apply to only $\sigma_{xy}(H)$ since the two-band model fittings can be already overparameterized, as discussed above. Even when we fitted the $\sigma_{xy}(H)$ and $\sigma_{xx}(H)$ curves simultaneously by adding an electron band to Eq.1 and Eq.2, the derived densities ($n_1$, $n_2$ and $n_3$) and mobilities ($\mu_1$, $\mu_2$ and $\mu_3$) still depend on parameter initialization.

On the other hand, if a hole band is added by using $\sigma_{xy}(H) = e\left(\frac{n_1\mu_1^2}{1+\mu_1^2 H^2} - \frac{n_2\mu_2^2}{1+\mu_2^2 H^2} + \frac{n_3\mu_3^2}{1+\mu_3^2 H^2}\right)H$ and $\sigma_{xx}(H) = \sum_i \frac{en_i\mu_i}{1+\mu_i^2 H^2}$ with $i = 1,2,3$, the fits can produce repeatable values of the densities and mobilities. However, the disparities between the fit and the experimental curves can become even larger for $\sigma_{xy}(H)$ while those for $\sigma_{xx}(H)$ are indeed decreased, as shown in Fig.S4. That said, fittings of the magnetoconductivity data, as demonstrated in both the two-band and three-band model analysis, is not a reliable method to quantitively determine the carrier density and mobility in a multi-band semimetal. In other words, anomalies in the temperature dependence of the derived carrier density and mobility may not be associated with Lifshitz transitions.

To further demonstrate that the carrier density and mobility in ZrSiSe may have no anomalies in their temperature dependence, we conducted Kohler's rule analysis on the experimental data. As shown in Fig.5, the magnetoresistances $MR$s for $T \leq 180$ K obey Kohler's rule very well, consistent with that reported in Ref.31. The $MR \sim H/\rho_0$ relationship shows a similar behavior observed in other multiband semimetals [25,32], i.e., following a power law of $MR \sim (H/\rho_0)^\alpha$ with $\alpha = 2$ at low magnetic fields and deviating from it at higher field values. Since Kohler's rule is valid for systems with temperature insensitive carrier densities [25], the scaling behavior in Fig.5 suggests the absence of anomalies in the temperature dependence of the carrier density at $T \leq 180$ K. We note that the curve for $T = 180$ K shifts slightly to the right of that for $T = 120$ K. Such a deviation from Kohler's rule is caused by the small increase of the carrier density due to thermal activation, which becomes more pronounced at higher temperatures, as manifested by the curves for $T = 250$ K and 300 K in Fig.S4(a). It can be accounted for by a change in the thermal factor $n_T$ in the extended Kohler's rule $MR \sim H/(n_T\rho_0)$ [25] as shown in Fig.S4(b), where values $n_T = 1.03$, 1.11 and 1.18 are derived for $T = 180$ K, 250 K and 300 K, respectively by assuming $n_T = 1$ for $T = 120$ K (and below).

## IV. CONCLUSIONS

In summary, we have shown that anomalies in the temperature dependence of the carrier density and/or mobility derived from the analysis of magnetoconductivity using the conventional two-band model may not be a reliable method to detect Lifshitz transitions. We addressed the reported Lifshitz transitions at $T \geq 80$ K in ZrSiSe, which are inferred from the anomalies in the carrier density and mobility in the second band from the two-band model analysis on the Hall magnetoconductivities. We showed that transitions can be an artifact of the fitting, which is overparameterized and can output arbitrary values of the carrier density and mobility for the second band for $T > 80$ K. The unreliability of the two-band model fitting is also affirmed by the large disparities between the experimental longitudinal magnetoconductivity curves and those calculated with the densities and mobilities derived from fitting the longitudinal magnetoconductivity curves [Fig.2(b) and 2(d)] as well as the entirely different outcomes [Fig.S2(c) and S2(d) versus Fig.2(e) and 2(f)] when both the longitudinal and Hall magnetoconductivities are fitted simultaneously. It is further attested by the scaling behavior of the magnetoresistance following the (extended) Kohler's rule, which suggests the absence of anomalies in the carrier densities. Our work demonstrates that cautions need be taken in interpreting the temperature dependences of the carrier density and mobility derived from the two-band model analysis on magnetoresistivities of a multi-band system.

**ACKNOWLEDGMENTS**

Magneto-transport measurements and data analysis were supported by the U.S. Department of Energy, Office of Science, Basic Energy Sciences, Materials Sciences and Engineering. Crystal growth at Penn State was supported by the US Department of Energy under grants DE-SC0019068. S. E. P. and Z. L. X. received support by National Science Foundation (Grant No.

DMR-1901843). Work performed at the Center for Nanoscale Materials, a U.S. Department of Energy Office of Science User Facility, was supported by the U.S. DOE, Office of Basic Energy Sciences, under Contract No. DE-AC02-06CH11357.

[*]Correspondence to: xiao@anl.gov

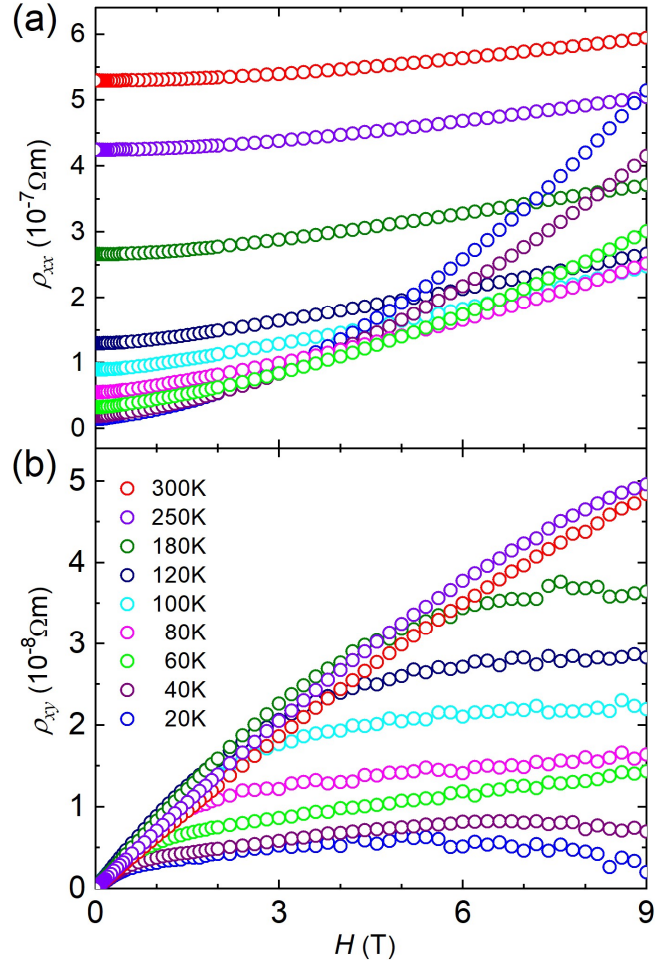

**Fig.1.** Magnetic field dependence of (a) the longitudinal magnetoresistivity $\rho_{xx}(H)$ and (b) the Hall magnetoresistivity $\rho_{xy}(H)$ obtained at various temperatures (For clarity, curves for $T < 20$ K are omitted since they are nearly indistinguishable to that for $T = 20$ K). The data were taken at $H//c$ and the current flows in the $ab$ plane. The sample's residual resistivity ratio $RRR = \rho_0(300K)/\rho_0(3K)$ is 39. Symbols for data taken at various temperatures are the same in both (a) and (b).

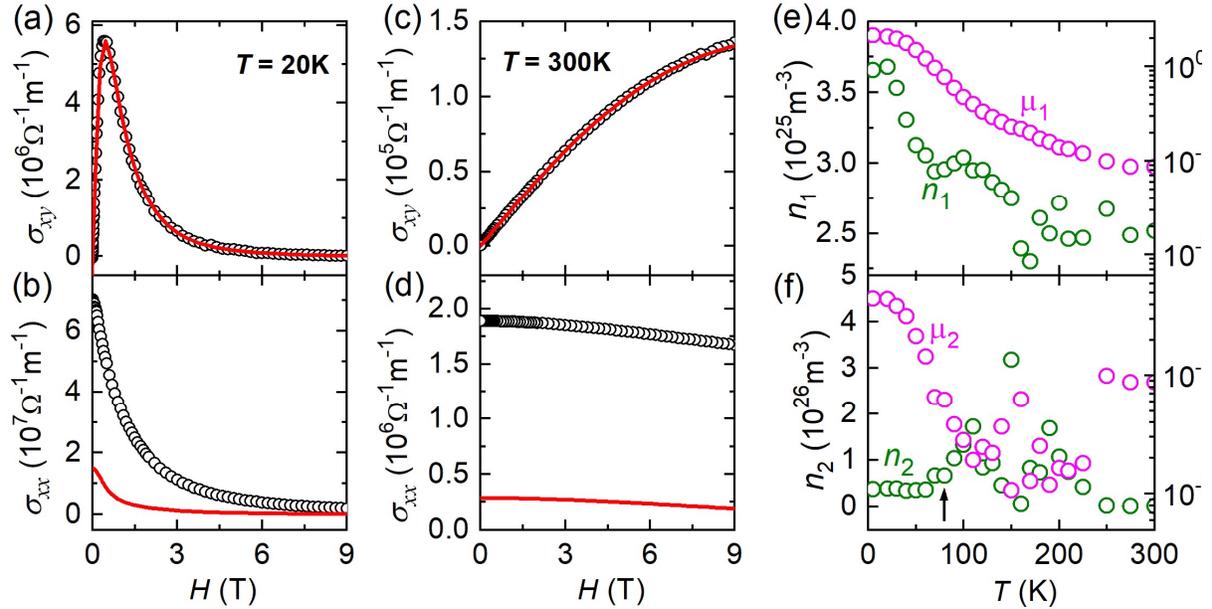

**Fig.2.** Two-band model analysis of Hall magnetoconductivities $\sigma_{xy}(H)$. (a) and (c) show examples of fits (red lines) of Eq.1 to experimental results (black circles) at $T = 20$ K and 300 K respectively. (b) and (d) compare the calculated longitudinal magnetoconductivities $\sigma_{xx}(H)$ (red lines) using the fitting parameters obtained in (a) and (c) with the experimental data (black circles). The large separation between the red lines and the black circles in (b) and (d) indicates that the parameters obtained in fitting $\sigma_{xy}(H)$ do not describe the $\sigma_{xx}(H)$ properly (see more discussions in the text). (e) and (f) show the temperature dependence of the derived fitting parameters for the first ($n_1$, $\mu_1$) and the second ($n_2$, $\mu_2$) bands, respectively. The fluctuation of the fitting parameters, particularly $n_2$ and $\mu_2$ in (f) at $T > 80$ K, is due to the negligible contribution $\sigma_{xy\_2}$ of the second band in Eq.1 to the total magnetoconductivity $\sigma_{xy}(H)$. The arrow in (f) marks the temperature of $T = 80$ K.

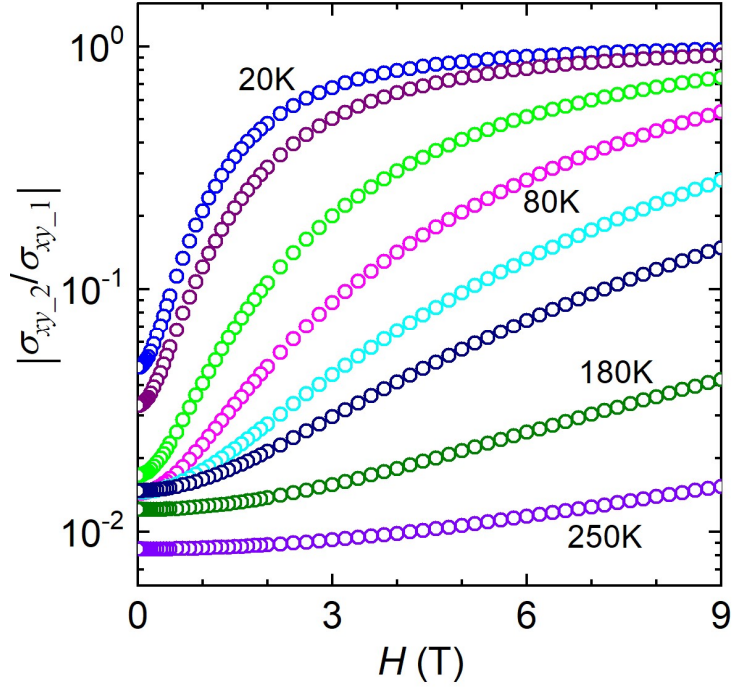

**Fig.3.** Magnetic field dependence of the ratio of the magnetoconductivity $\sigma_{xy\_2}$ of the second band to that ($\sigma_{xy\_1}$) of the first band in the two-band model analysis Eq.1, where $\sigma_{xy\_i} = en_i\mu_i^2 H/[1+(\mu_i H)^2]$ with $i = 1$ and 2. The symbols are the same as those in Fig.1.

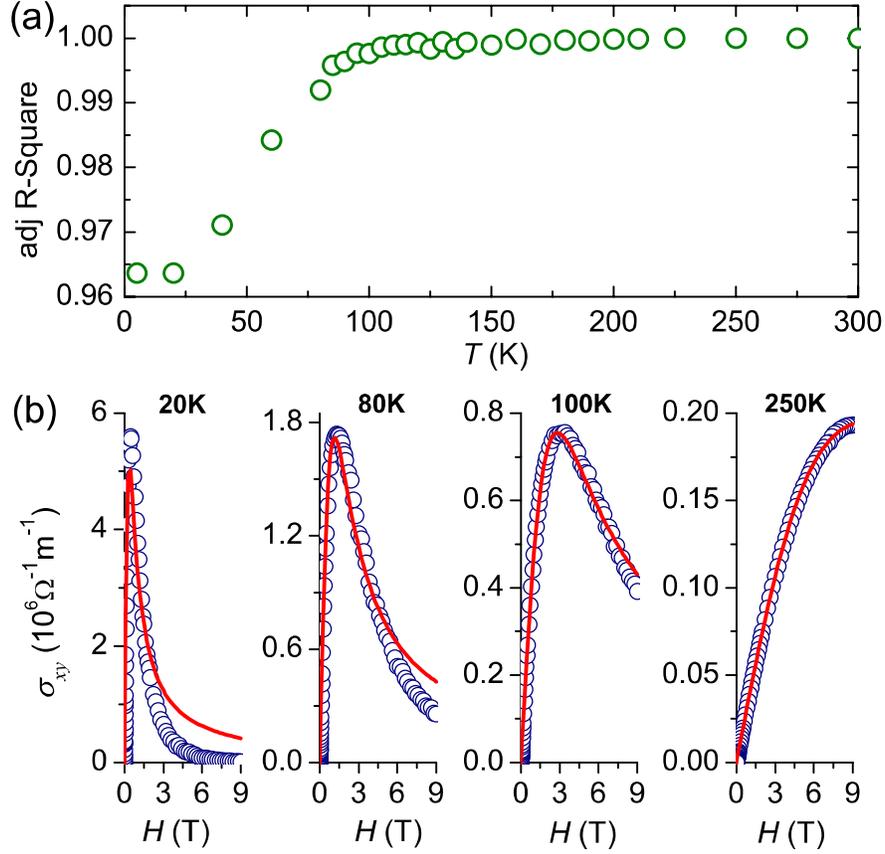

**Fig.4.** (a) Temperature dependence of the Adj. R-square obtained in one-band fittings of the experimental $\sigma_{xy}$, i.e., by setting $n_2$ and $\mu_2$ in Eq.1 to be 0. Here, Adj. R-square is a modified version of R-square that is also known as the coefficient of determination (COD). It is a statistical measure to qualify the linear regression and is always less than 1. The larger the Adj. R-square, the better the fit. When the Adj. R-square approaches 1, the fitted line explains all the variability of the response data around its mean [29]. (b) Examples showing the quality of the one-band fittings at various temperatures. The blue circles represent experimental $\sigma_{xy}$ while red lines are the fits.

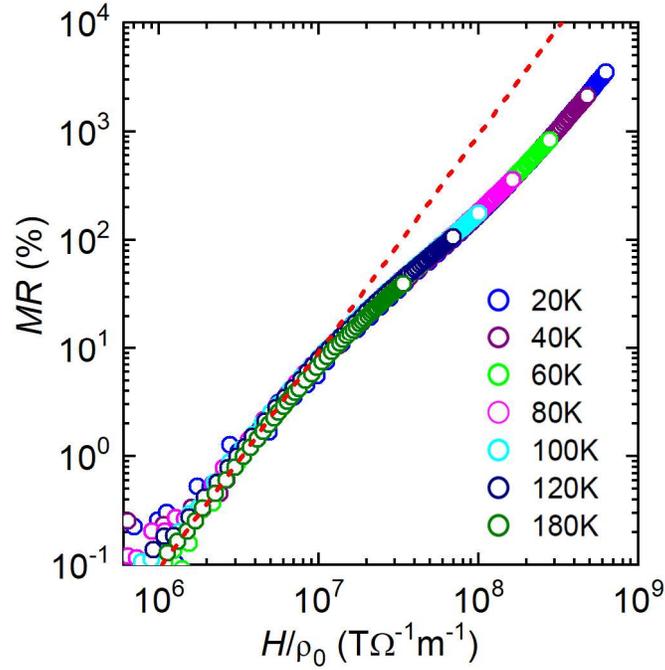

**Fig.5.** Kohler's rule plots of longitudinal magnetoresistivity $\rho_{xx}(H)$ curves in Fig.1, implying the absence of a temperature-driven Lifshitz transition at $T$ = 80 K -180K. The symbols are the same as those in Fig.1. For clarity, only data up to $T$ = 180 K are presented. Kohler's rule plots of all data from Fig.1 and discussions on thermal induced change in the carrier density are presented in the supplement. The red dotted line represents $MR \sim (H/\rho_0)^2$.

Supplementary Information:

**Unreliability of two-band model analysis of magnetoresistivities in unveiling temperature-driven Lifshitz transition**

Jing Xu[1,2], Yu Wang[3], Samuel E. Pate[1,4], Yanglin Zhu[3], Zhiqiang Mao[3], Xufeng Zhang[2], Xiuquan Zhou[1], Ulrich Welp[1], Wai-Kwong Kwok[1], Duck Young Chung[1], Mercouri G. Kanatzidis[1,5], and Zhi-Li Xiao[1,2,*]

[1]*Materials Science Division, Argonne National Laboratory, Argonne, Illinois 60439, USA*

[2]*Center for Nanoscale Materials, Argonne National Laboratory, Argonne, Illinois 60439, USA*

[3]*Department of Physics, Pennsylvania State University, University Park, Pennsylvania 16802, USA*

[4]*Department of Physics, Northern Illinois University, DeKalb, Illinois 60115, USA*

[5]*Department of Chemistry, Northwestern University, Evanston, Illinois 60208, USA*

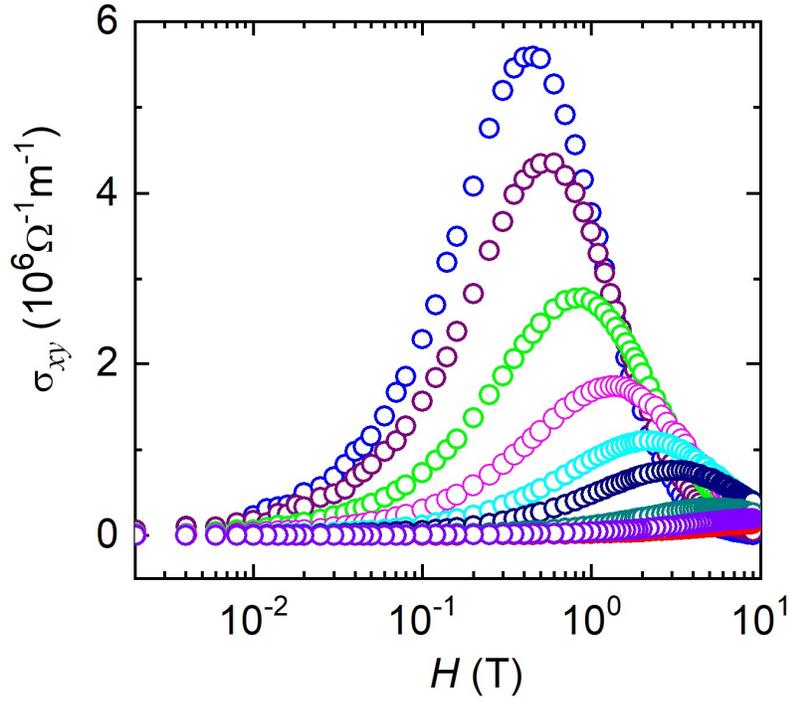

**Fig.S1.** Hall magnetoconductivities $\sigma_{xy}(H)$ converted from $\rho_{xy}(H)$ and $\rho_{xx}(H)$ at various temperatures in Fig.1. The symbols are the same as those in Fig.1.

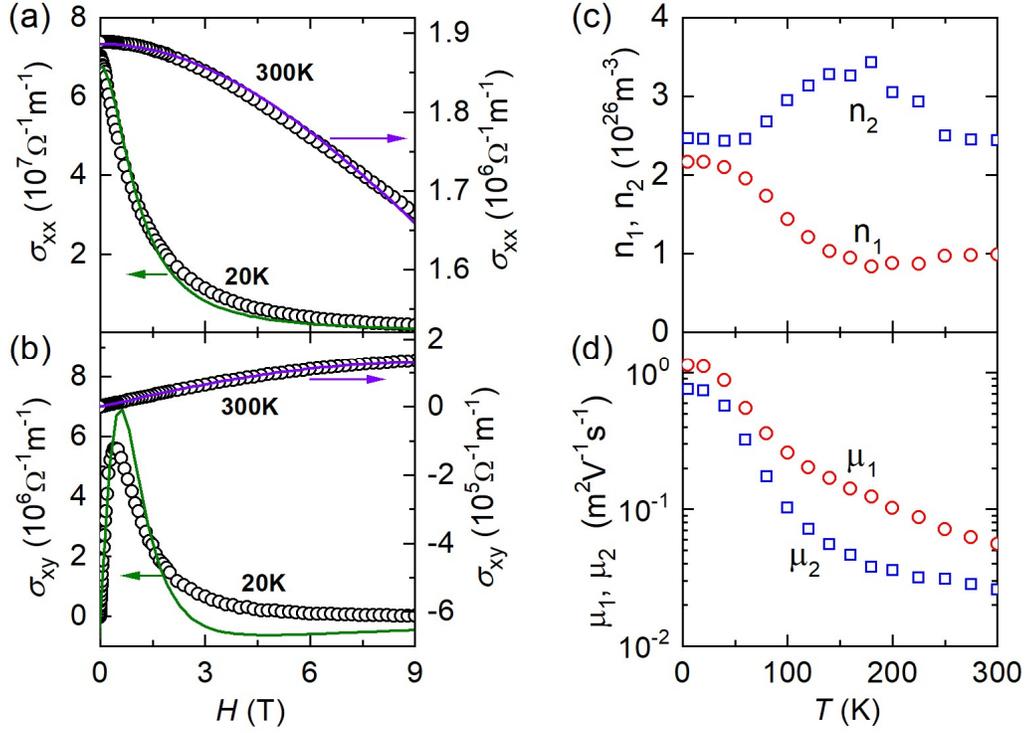

**Fig.S2.** Two-band model analysis of magnetoconductivities with fittings of both $\sigma_{xx}(H)$ and $\sigma_{xy}(H)$ simultaneously. (a) and (b) Fits (olive and purple lines) to experimental results (black circles) of $\sigma_{xx}(H)$ and $\sigma_{xy}(H)$, respectively. Data taken at $T = 20$ K and 300 K are shown as examples. (c) and (d) Temperature dependences of the derived carrier densities ($n_1$, $n_2$) and mobilities ($\mu_1$, $\mu_2$), respectively.

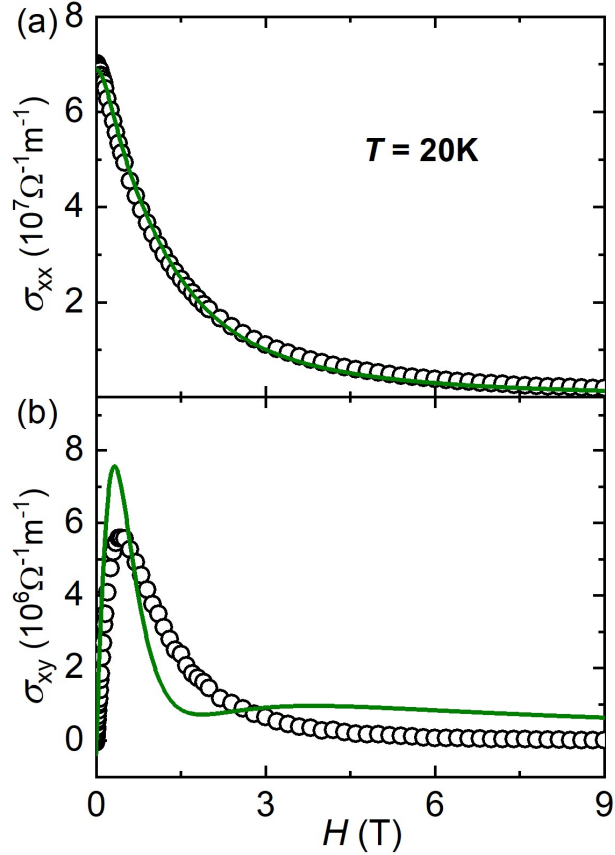

**Fig.S3.** Three-band model analysis of magnetoconductivities with fittings of both $\sigma_{xx}(H)$ and $\sigma_{xy}(H)$ simultaneously. (a) and (b) Fits (olive lines) to experimental results (black circles) of $\sigma_{xx}(H)$ and $\sigma_{xy}(H)$, respectively. Data taken at $T = 20$ K. The derived parameters are $n_1 = 2.12103\times10^{26}$ m$^{-3}$, $n_2 = 2.3234\times10^{26}$ m$^{-3}$, $n_3 = 0.59548\times10^{26}$ m$^{-3}$ and $\mu_1 = 0.53584$ m$^2$V$^{-1}$s$^{-1}$, $\mu_2 = 0.75413$ m$^2$V$^{-1}$s$^{-1}$, $\mu_3 = 2.37843$ m$^2$V$^{-1}$s$^{-1}$. Among the three bands, the first and third ones are hole bands and the second one is an electron band (see discussions in the text).

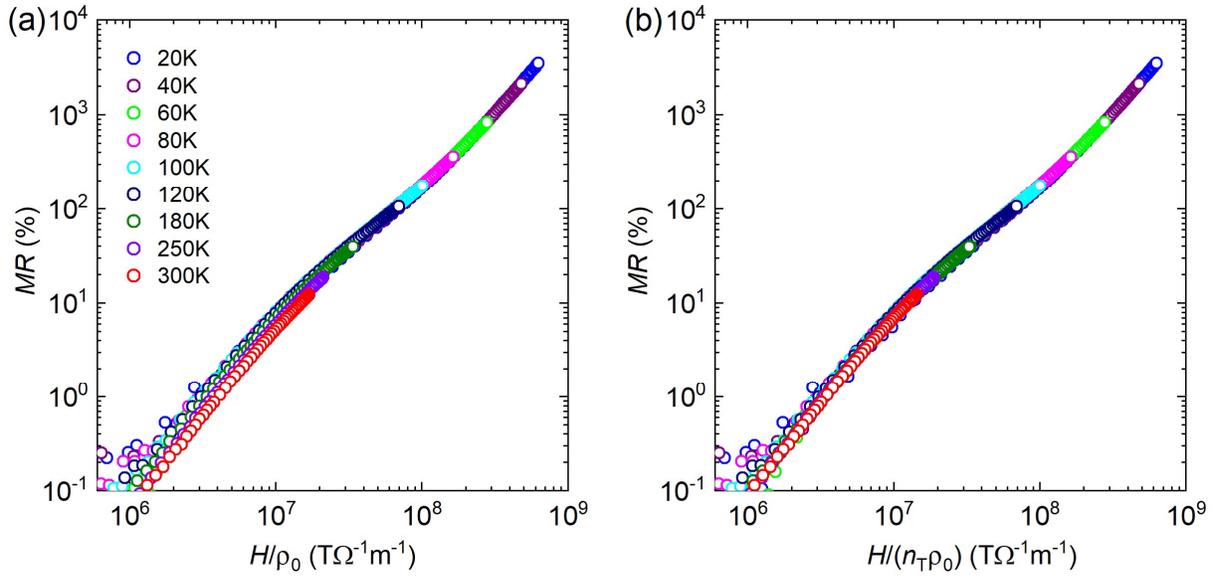

**Fig.S4.** (a) Kohler's rule plot and (b) Extended Kohler's rule plot of the magnetoresistivities data in Fig.1(a). We assumed $n_T = 1$ for $T = 120$ K (and below) and obtained $n_T = 1.03$, $1.12$, and $1.18$ for $T = 180$ K, $250$ K, and $300$ K, respectively.